\documentclass[showpacs,showkeys,preprintnumbers,amsmath,amssymb,latexsym]{revtex4}

\usepackage{graphicx}

\def\beq{\begin{equation}}
\def\eeq{\end{equation}}

\begin{document}

\title{The effective geometry of the $n=1$ uniformly rotating self-gravitating polytrope}

\author{D. Bini}
\affiliation{
Istituto per le Applicazioni del Calcolo ``M. Picone,'' CNR, I-00185 Rome, Italy and\\
ICRA, University of Rome ``La Sapienza,'' I-00185 Rome, Italy}

\author{C. Cherubini}
\affiliation{Nonlinear Physics and Mathematical Modeling Lab,
Engineering Faculty, University Campus Bio-Medico,
I-00128 Rome, Italy and\\
ICRA, University of Rome ``La Sapienza,'' I-00185 Rome, Italy}

\author{S. Filippi}
\affiliation{Nonlinear Physics and Mathematical Modeling Lab,
Engineering Faculty, University Campus Bio-Medico,
I-00128 Rome, Italy and\\
ICRA, University of Rome ``La Sapienza,'' I-00185 Rome, Italy}

\author{A. Geralico}
  \affiliation{Physics Department and
ICRA, University of Rome ``La Sapienza,'' I-00185 Rome, Italy}

\begin{abstract}
The \lq\lq effective geometry" formalism is used  to study the
perturbations of a perfect barotropic Newtonian self-gravitating
rotating and compressible fluid coupled with gravitational
backreaction. The case of a uniformly rotating polytrope
with index $n=1$ is investigated, due to its analytical tractability.
Special attention is devoted to the geometrical
properties of the underlying background acoustic metric, focusing
in particular on null geodesics as well as on the analog
light cone structure.
\end{abstract}



\maketitle

\section{Introduction}

The perturbative study of an equilibrium configuration of a given
perfect barotropic and  irrotational Newtonian fluid using an ``effective spacetime metric'' has been
formerly addressed by Unhru \cite{Unruh94} and Visser
\cite{visser}. Shortly, the equations satisfied by the perturbed
quantities are unified in a linear second order hyperbolic
equation with non-constant coefficients for the velocity potential
only, while other quantities like density and pressure follow directly. The theory is formally
equivalent to the dynamics of a massless scalar field on a
pseudo-Euclidean four dimensional Riemannian manifold, leading in
this way to an \lq\lq effective gravity.'' These pioneering studies started
then a series of further publications in the new field of \lq\lq
analog geometries in condensed matter," which generalized the
initial hypotheses in a number of contexts (see Refs.
\cite{BARCELO,NOVELLO,SCHULTZ} for relatively recent overviews on
this subject).

In a recent paper \cite{bcfprd} this theoretical framework has
been extended to the study of irrotational perturbations of
spherically symmetric self-gravitating polytropic configurations
for different values of the polytropic index $n$, considering  the
coupling of hydrodynamics and gravitation. Polytropic systems have
been chosen essentially because such a specific equation of state
plays an important role in galactic dynamics as well as in the
theory of stellar structure \cite{Tassoul, Binney}, although  the
formalism developed can be directly applied also to non-barotropic
equations of state. Furthermore, it is well known in the
literature that polytropes with index $n=0$ and $n=1.5$ describe a
homogeneous liquid state and a monoatomic gas in adiabatic
equilibrium respectively, whereas the limiting case $n\to \infty$
represents an isothermal perfect gas \cite{OSTRIKER}. While in the
previous studies on effective gravity models (or \lq\lq analog
models'') the contribution of gravitational field was assumed to
be externally fixed (i.e. without backreaction), in the
generalized framework presented in \cite{bcfprd} the effect of
gravitational backreaction  was taken into account. The problem
presents in this case two different time-scales: the acoustic one,
governed by hydrodynamics and allowing for acoustic waves
propagating at finite speed, and  the gravitational one, which in
a Newtonian picture allows for waves propagating at infinite speed
(General Relativity is required to correct this pathology).

In the present paper we extend the results of Ref. \cite{bcfprd} to the
case of rotating polytropes. To this aim we need an
analytical background solution for the Poisson-Euler's nonlinear
system of equations.  Unfortunately, in contrast with the spherical
non-rotating case, in the uniformly rotating situation the
polytropic fluid (star or galaxy) having also nonzero vorticity has analytic solutions only in
the case $n=1$, found by Williams in 1988 \cite{williams}.
Solutions for other
polytropic indices and for other more complicated rotation fields
can be obtained by numerical techniques
only (see Ref. \cite{oredth} for a comprehensive discussion of this
point). For these reasons, using this analytical solution as
background field, we write down the coupled system
of equations for the acoustic perturbations. Here the situation is
more complicated with respect to the non-rotating case developed
in Ref. \cite{bcfprd}, because if the flow has non-zero vorticity
 in the background flow this couples with the
perturbations and generates vorticity in the fluctuations. A
mathematical gauge-invariant formulation of the problem based on
Clebsch potentials was presented in Ref. \cite{ROTA} and is
extended here in order to account for the gravitational
backreaction. A straightforward calculation allows us to cast
this acoustic problem in an analogous general relativistic form,
defining an acoustic metric. The geometric quantities of the
metric associated with the uniformly rotating $n=1$ case are
studied in detail, including curvature invariants and analog light
cone structure. In particular, analyzing specific classes of
geodesics, we get enough information about the behavior of
perturbations, otherwise requiring nontrivial numerical techniques.

The article is organized as follows. After this introductory
section we present the general theory of self-gravitating
fluid/gaseous masses both at exact and perturbative level,
generalizing the existing Clebsch theory. We then describe Williams'
analytical solution for the uniformly rotating $n=1$ polytrope
together with its associated acoustic metric. Finally, we discuss
the physical implications of our study and its possible future
extensions.

\section{Self-gravitating barotropes}

\subsection{General equations}

Let us shortly recall the theory underlying classical
self-gravitating masses.
Although  the exact formulation of the problem can be found in many astrophysics textbook \cite{Tassoul,Binney},
we shall present here a less known approach based on the
Clebsch potential formalism \cite{LAMB,kambe}, further extended by
including the effects of a gravitational potential.
Following Ref. \cite{ROTA}, the starting
point of this approach is to define three potentials, $\phi$,
$\beta$ and $\gamma$, and the action
\beq
S=\int dtd^3x \left\{ -\frac 12 \rho
v^2 - \phi[\dot \rho+\nabla\cdot (\rho \vec v)]+ \rho\beta [\dot
\gamma +(\vec v\cdot \nabla) \gamma]+u(\rho)+\frac{1}{8\pi
G}(\nabla\Phi)^2+\rho\Phi  \right\}\,,
\label{EQ:action}
\eeq
where $\rho$ is the fluid mass-density, $\vec v$ the velocity,
$\Phi$ the gravitational potential and $u(\rho)$ the internal
energy density, an overdot denoting differentiation with respect to time.
Requiring stationarity for $S$ and varying for $\vec
v$ one obtains the Clebsch representation of the velocity field, i.e. $\vec v=\nabla \phi+\beta \nabla
\gamma$, which allows for flows with nonzero vorticity $\vec\omega\equiv\nabla\times\vec v =\nabla \beta \times\nabla
\gamma$.
We point out here a difference in sign with respect to the definition of the velocity potential adopted in Ref. \cite{bcfprd}.
Our action then becomes
\beq
S=\int dt
d^3x\left\{\frac12\rho(\nabla\phi+\beta\nabla\gamma)^2+\rho(\dot\phi+\beta\dot\gamma)+u(\rho)+\frac{1}{8\pi
G}(\nabla\Phi)^2+\rho\Phi\right\}
\equiv\int dt d^3x {\cal L}\,.
\label{BELLA}
\eeq
The equations of motion can now be derived by performing variations with respect to the remaining variables in Eq. (\ref{BELLA}) according to \cite{Greiner,Greiner2}
\beq
\frac{\partial {\cal L}}{\partial \Psi_A}-\frac{\partial}{\partial
x^i}\frac{\partial {\cal L}}{\partial (\partial\Psi_A/\partial
x^i)}-\frac{\partial}{\partial t}\frac{\partial {\cal L}}{\partial\dot\Psi_A}=0\,,\qquad A=1\ldots5\,,\quad i=1\ldots3\,,
\label{ELEQN}
\eeq
for a set of scalar fields $\Psi_A$ (in our case $\Psi_A=(\phi,\beta,\gamma,\rho,\Phi)$).
We thus find
\begin{eqnarray}
&&\dot\rho+\nabla\cdot(\rho\vec v)=0\quad\quad\quad\quad\quad\quad
({\rm varied}\!:\delta
\phi)\nonumber\\
&& \dot \gamma+(\vec v \cdot \nabla)\gamma=0
\quad\quad\quad\quad\quad\quad ({\rm varied}\!:\delta
\beta)\nonumber\\
&& \dot \beta+(\vec v \cdot \nabla)\beta=0
\quad\quad\quad\quad\quad\quad ({\rm varied}\!:\delta
\gamma)\nonumber\\
&& \frac12v^2+\dot\phi+\beta\dot\gamma+h+\Phi=0  \,\,\,\quad ({\rm
varied}\!:\delta
\rho)\nonumber\\
&&\nabla^2\Phi=4\pi G\rho  \,\,\quad\quad\quad\quad\quad\quad\quad
({\rm varied}\!:\delta \Phi)
\label{SYSTEMA}
\end{eqnarray}
where $h=du/d\rho$ is the specific entalpy. We point out that both
$\beta$ and $\gamma$ are advected with the fluid motion. The first
equation of Eq. (\ref{SYSTEMA}) is the continuity equation,
whereas the second, third and fourth equations reproduce Euler's
equations. The last equation is the Poisson's equation describing
gravitation inside the fluid. Outside the fluid mass the
gravitational potential satisfies instead Laplace's equation, i.e.
\begin{equation}
\nabla^2 \Phi_{\rm ext} =0\,. \label{Gravout}
\end{equation}
Both the inner and outer Newtonian potentials and their normal
gradients must be matched at the configuration's boundary.

\subsection{Linear perturbations}

Let us study the evolution of small fluctuations $(\rho_1,
\phi_1,\beta_1, \gamma_1, \Phi_1)$ superimposed on  a background
flow $(\rho_0,\phi_0, \beta_0,\gamma_0,\Phi_0)$, which we will
assume to be iso-entropic but neither steady nor incompressible.
The linearization procedure consist in expanding the action (\ref{BELLA}) up to quadratic order in the
fluctuations, so that $S=S_0+S_1+S_2+\cdots$. The zeroth order
variables obey the background field equations, whereas the first order action
$S_1$, which contains terms  linear in the fluctuations, is identically
zero as an easy check can show.
The quadratic term turns out to be
\beq
S_2=
\int dtd^3x\left\{\frac 12 \rho_0 v_1^2 +\rho_1 \vec v_0\cdot \vec
v_1 +\rho_1(\dot \phi_1+ \beta_0\dot\gamma_1+\beta_1\dot \gamma_0)
+\rho_0\beta_1 \dot \gamma_1 +\frac 12 \frac{c^2}{\rho_0}
\rho_1^2+\rho_1\Phi_1+\frac{1}{8\pi G}(\nabla\Phi_1)^2\right\}\,,
\label{EQS2}
\eeq
where we have introduced the first order velocity field
\beq
\label{v1def}
\vec v_1=\nabla\phi_1
+\beta_1 \nabla\gamma_0 + \beta_0 \nabla\gamma_1\,,
\eeq
and the background local speed of sound $c$ defined by
\beq
\label{cdef}
c^2\equiv\left(\frac{dp}{d\rho}\right)_0= \rho_0\,\left(\frac{d^2
u}{d\rho^2}\right)_0\,.
\eeq
The latter relation follows from the iso-entropic properties of the flow.
In fact, Taylor expanding the internal energy density gives
\beq
u=\rho e\simeq(\rho e)_0+\left[\frac{d(\rho
e)}{d\rho}\right]_0\rho_1+\frac{1}{2}\left[\frac{d^2(\rho
e)}{d\rho^2}\right]_0\rho_1^2\,,
\eeq
where $e$ represents the internal energy, the last term entering the expression for $S_2$.
The iso-entropic condition
$de=(p/\rho^2)d\rho$ (see Ref. \cite{kambe} p. 40
or Ref. \cite{LLflmech} p. 255 for further details) then implies
\beq
\frac{du}{d\rho}=
e+\frac{p}{\rho}\quad\to\quad
\frac{d^2
u}{d\rho^2}=
\frac{1}{\rho}\frac{dp}{d\rho}
\equiv \frac{c^2}{\rho}\,,
\eeq
finally leading to Eq. (\ref{cdef}).

The Clebsch decomposition of the velocity field is not gauge invariant, since the potentials are not uniquely determined.
In order to deal with physical variables, it is convenient to express the first order velocity fluctuation (\ref{v1def}) into two gauge-invariant parts
\beq
\label{v1def2}
\vec v_1=
\nabla\psi_1+\vec \xi_1\,,
\eeq
where the scalar field  $\psi_1=\phi_1+\beta_0\gamma_1$ can be identified with the acoustic degree of freedom and the vector field $\vec \xi_1=\beta_1\nabla\gamma_0-\gamma_1\nabla\beta_0$ with a partial hybridization of the sound with other modes \cite{ROTA}.
The latter has been shown in Ref. \cite{ROTA} to be a small correction to the potential flow and equal to $\vec \xi_1=\vec x_1\times\omega_0$, where $\vec x_1$ represents the particle displacement caused by the sound wave.

By varying now the action (\ref{EQS2}) with respect to $\rho_1$ and using the
background equations  we obtain
\beq
\label{PEZZO1}
\rho_1=-\frac{\rho_0}{c^2}\left[\frac{D^{(0)}\psi_1}{dt}+\Phi_1\right]\,,
\eeq
where we have defined
\beq
\frac{D^{(0)}}{dt}=\partial_t+\vec
v_0\cdot\nabla \eeq
as the background convective derivative.
The linearization of the continuity equation then gives
\beq
\partial_t\rho_1+\vec v_0\cdot \nabla\rho_1+\rho_1\nabla\cdot \vec
v_0+\nabla\cdot(\rho_0 \vec v_1)=0\,.
\label{CONT1}
\eeq
Next substituting Eq. (\ref{PEZZO1}) into Eq. (\ref{CONT1}) and using the background continuity
equation together with the definition (\ref{v1def2}) of the first order velocity field, we finally obtain after some algebra  the wave equation
\beq
\left(\frac{D^{(0)}}{d
t}\right)\frac{1}{c^2}\left(\frac{D^{(0)}}{d
t}\right)\psi_1-\frac{1}{\rho_0}\nabla
\cdot\left(\rho_0\nabla\psi_1\right)=\frac{1}{\rho_0}\nabla
\cdot\left(\rho_0\vec \xi_1\right)-\left(\frac{D^{(0)}}{d
t}\right)\left(\frac{\Phi_1}{c^2}\right)\,.
\label{PRIMADI}
\eeq

The equation determining the time evolution of the quantity $\vec \xi_1$ can be worked out as
follows. We first observe that both $\beta$ and $\gamma$ are
convectively conserved, so that we have
\beq
\frac{D^{(0)}}{dt}\beta_0=0\,,
\label{BETA0}
\eeq
and
\beq
\partial_t \beta_1 +
(\vec v_0\cdot \nabla) \beta_1 + (\vec v_1 \cdot \nabla)
\beta_0=0\,.
\eeq
Taking then the gradient of Eq. (\ref{BETA0}) leads to
\beq
\left(\frac{D^{(0)}}{dt}\right) \nabla_i
\beta_0 = - (\nabla_i v_{0j}) \nabla_j \beta_0\,,
\eeq
where summation over repeated indices is understood as a standard convention.
Analog equations can be written for the potential $\gamma_0$ too.
Recalling now the definition of $\vec\xi_1$, these relations can be combined to give
\beq
\left(\frac{D^{(0)}}{dt}\right) \xi_{1i} =
-\nabla_j\psi_1(\nabla_j v_{0i} -
      \nabla_i v_{0j})
           -\xi_{1j}(\nabla_j v_{0i})\,,
\label{CIRCLE}
\eeq
which in vector notation becomes
\beq
\frac {D^{(0)} \vec \xi_1}{dt} = \nabla \psi_1 \times
\vec \omega_0 - (\vec \xi_1\cdot \nabla)\vec v_0\,.
\label{PERXI}
\eeq

As concerns the gravitational part, by performing the variation of Eq. (\ref{EQS2}) with
respect to $\Phi_1$ and taking into account Eq. (\ref{PEZZO1}) we find
\beq
\label{Grav2}
\nabla^2\Phi_1=4\pi G\rho_1
=-\frac{4\pi G\rho_0}{c^2}\left[\frac{D^{(0)}}{d
t}\psi_1+\Phi_1\right]\,.
\eeq
Outside the configuration instead Eq. (\ref{Gravout}) still holds.

The coupled system of equations (\ref{PRIMADI}), (\ref{PERXI}) and (\ref{Grav2}) thus form a complete closed set of equations describing the first order fluctuation about the background flow in terms of gauge-invariant quantities only.

\subsection{The effective geometry}

Following Visser \cite{visser} let us form the symmetric $4\times4$ matrix
\beq
f^{00}=-\frac{\rho_0}{c^2}\ , \quad
f^{0i}=-\frac{\rho_0v_0^i}{c^2}\ ,\quad
f^{ij}=\rho_0\left(\delta^{ij}-\frac{v_0^iv_0^i}{c^2}\right)\,,
\eeq
where Roman indices run from $1$ to $3$.
Next define the symmetric tensor $g^{\mu\nu}$ by $\sqrt{-g}\, g^{\mu\nu} =
f^{\mu\nu}$, implying that $\sqrt{-g} = \rho_0^2/ c$, Greek indices running from $0$ to $3$.
We finally get the acoustic line element
$ds^2=g_{\mu\nu}dx^\mu dx^\nu$ with metric tensor and its inverse
given by
\begin{equation}
\label{acousticmetric}
g_{\mu\nu} \equiv \frac{\rho_0}{c} \left[
\begin{array}{ccc}
-(c^2-v_0^2)&\vdots&-v_0^j\\
               \cdots\cdots\cdots\cdots&\cdot&\cdots\cdots\\
           -v_0^i&\vdots&\delta_{ij}\\
\end{array}
\right]
\,,\qquad
g^{\mu\nu}\equiv \frac1{\rho_0 c} \left[
\begin{array}{ccc}
-1&\vdots&-v_0^j\\
               \cdots\cdots&\cdot&\cdots\cdots\cdots\cdots\\
           -v_0^i&\vdots&(c^2 \delta^{ij} - v_0^i v_0^j )\\
\end{array}
\right]\,,
\end{equation}
respectively.
Using then coordinates $x^\mu \equiv (t, x^i)$ the wave equation (\ref{PRIMADI})
can  be rewritten in the more suggestive form
\beq
\frac 1{\sqrt{-g}}
\partial_\mu \left({\sqrt{-g}}g^{\mu\nu}\partial_\nu \psi_1\right)=\frac{1}{\sqrt{-g}}\left[\nabla \cdot\left(\rho_0\vec
\xi_1\right)-\rho_0\left(\frac{D^{(0)}}{d
t}\right)\left(\frac{\Phi_1}{c^2}\right)\right]\,.
\label{GEOMETRO}
\eeq

Summarizing, the final system of equations results in
\begin{eqnarray}
\label{FINALSRISCR}
\nabla^\mu\nabla_\mu\psi_1&=&
\frac 1{\sqrt{-g}}
\left[\nabla \cdot\left(\rho_0\vec
\xi_1\right)-\rho_0\left(\frac{D^{(0)}}{d
t}\right)\left(\frac{\Phi_1}{c^2}\right)\right]
\,,\nonumber\\
\frac {D^{(0)} \vec \xi_1}{dt} &=& \nabla \psi_1 \times
\vec \omega_0 - (\vec \xi_1\cdot \nabla)\vec v_0
\,,\nonumber\\
 \left[\nabla^2+k_J^2\right] \Phi_1&=&-k_J^2
\frac{D^{(0)}}{dt}\psi_1\,,
\end{eqnarray}
which mixes the hydrodynamical (with finite speed) and
gravitational (instantaneous)  problems  through first order time
and space partial derivatives of the fields. Here $\nabla_\mu$ in
the first equation denotes the covariant derivative with respect
to the acoustic metric $g_{\mu\nu}$, whereas $\nabla^2$ in the
third equation is the standard Laplace operator of Euclidean space
in three dimensions. Finally, the quantity  $k_J=\sqrt{4\pi
G\rho_0/c^2}$ is the magnitude of a generalized Jeans' wavevector.

\section{Perturbations of the $n=1$ rotating polytrope}

Let us apply the formalism developed in the previous section to the case of a uniformly
rotating polytropic star with index $n=1$, whose analytical solution has been found by Williams \cite{williams}.

Such a specific choice is not motivated simply by important
mathematical reasons (see as an example \cite{Taniguchi} for a
recent discussion on the simplifications occurring in finding
analytical solutions for  astrophysical binary systems when $n=1$
is chosen), because in this case the nonlinear gravitational and
hydrostatic problem collapses into a linear one. Realistic
equations of state for neutron star matter in fact seem to require
$n\simeq1$ \cite{Weber, Lattimer,Stergioulas}. Furthermore,
theoretical studies for the spherical configurations
\cite{Chandra} show that in the case $n=1$ the radius of the
configuration is independent of the central density, so that for a
configuration in convective polytropic equilibrium the radius will
depend on this polytropic temperature only. Finally the value of
$n$ chosen fits well with the estimates of stability  rotating
polytropic stars \cite{Tassoul,oredth,Chandra,james,Imamura}. We
shall assume throughout this section that all physical quantities
refer to background solution of the exact nonlinear problem.

\subsection{Equilibrium configurations}

The basic equations  governing the hydrostatic
equilibrium of a self-gravitating axially symmetric fluid rotating
with uniform angular velocity $\Omega$ are given by
\begin{eqnarray}
\label{equileqns}
\nabla p &=& -\rho \nabla \left[\Phi-\frac12\Omega^2r^2\sin^2\theta\right]\,, \nonumber\\
\nabla^2 \Phi&=&4\pi G\rho\,,
\end{eqnarray}
the gravitational potential outside the fluid being governed by the Laplace equation
\begin{equation}
\nabla^2 \Phi_{\rm ext} =0\,.
\end{equation}
The velocity field in this case is in fact given by $\vec
v=\Omega(-y\partial_x+x\partial_y)=\Omega
r\sin\theta\partial_\phi$, implying that the fluid has nonzero vorticity $\vec\omega=2\Omega\partial_z$.

Let the fluid be described by a
polytropic equation of state (see Ref. \cite{Chandra} for
details), i.e. \beq p=K\rho^{1+1/n}\,, \eeq with the inverse
relation \beq \rho=\left(\frac{p}{K}\right)^{n/(n+1)}\,, \eeq so
that sound speed in this case results in \beq
c=\left(\frac{\partial \rho}{\partial
p}\right)^{-1/2}=K^{1/2}\left(1+\frac{1}{n}\right)^{1/2}\rho^{1/2n}\,.
\eeq

In order to simplify the integration of the equations let us
introduce the so called Lane-Emden parametrization
\cite{Tassoul,Chandra1}, i.e. $\rho=\rho_c\Theta^n$, $r=\alpha
\xi$ with
$\alpha=(4\pi G)^{-1/2}(n+1)^{1/2}K^{1/2}\rho_c^{(1-n)/2n}$,
 $\rho_c$ denoting the density at the center of the fluid
configuration. The first of Eqs. (\ref{equileqns}) can thus be
integrated yielding \beq
(n+1)K\rho_c^{1/n}\Theta=-\Phi+\frac12\Omega^2r^2\sin^2\theta\,,
\eeq up to a constant term which, in turn, can be re-absorbed in
the definition of the gravitational potential. After a suitably
rescaling of the gravitational potential too  we get the following
algebraic relation \beq \label{Thetachi}
\Theta=-\chi+\frac14\beta\xi^2(1-\mu^2)\,, \eeq where
$\mu=\cos\theta$, $\beta=\Omega^2/(2\pi G\rho_c)$ and
$\chi=\Phi/[(n+1)K\rho_c^{1/n}]$. The second  of Eqs.
(\ref{equileqns}) thus becomes the well known Lane-Emden equation
\beq \label{LANEE} \nabla^2\Theta=-\Theta^n+\beta\,, \eeq and in
presence of uniform rotation ($\beta \neq 0$) it admits an exact
analytic solutions for $n=1$ due to the linear nature of the PDE.
In the case of axial symmetry and for $n=1$ it becomes \beq
\label{LANEEneq1}
\partial_\xi(\xi^2\partial_\xi\Theta)+\partial_\mu[(1-\mu^2)\partial_\mu\Theta]=\xi^2(-\Theta+\beta)\ ,
\eeq
and it should be solved with the conditions $\Theta=1$ and
$\partial_\xi\Theta=0$ for $\xi=0$ as discussed in
Ref. \cite{Binney}.
This is an elliptic system subjected to the above initial conditions but with free boundary, since the surface of the star is not known a priori.
On the other hand, on the unknown star's surface both the external and
internal gravitational potential as well as their gradients
projected on the outgoing and ingoing normal directions
respectively must coincide.
Only after imposing all conditions on the unknown common boundary one will succeed in determining it, as explicitly shown by Williams \cite{williams}.

\begin{figure}
\typeout{*** EPS figure surface}
\begin{center}
\includegraphics[scale=0.5]{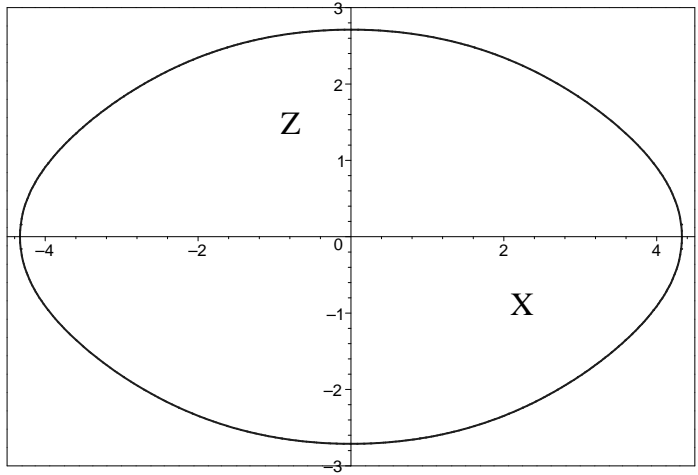}
\end{center}
\caption{The surface of the rotating polytrope is shown for
$\beta=8\times10^{-2}$ ($X=\xi\sin \theta$ and $Z=\xi\cos \theta$
are Cartesian-like coordinates). The polar and equatorial radii
are given by $\xi_-\approx2.7117$ and $\xi_+\approx4.3302$
respectively, in agreement  with both James ($\xi_-\approx2.7194$,
$\xi_+\approx4.3797$) \cite{james} and Williams
($\xi_-\approx2.7175$, $\xi_+\approx4.2733$) \cite{williams}
results. } \label{fig:surface}
\end{figure}

The solution of Eq. (\ref{LANEEneq1}) is given by \beq
\Theta=\beta+(1-\beta)\frac{\sin\xi}{\xi}+\sum_{l=2}^{\infty}\frac{b_l}{\sqrt{\xi}}J(l+1/2,\xi)P_l(\mu)\
, \eeq where $J(k,x)$ are Bessel polynomials and $P_l(x)$ are
Legendre polynomials, with $b_3=0,\ldots,b_{2k+1}=0,\ldots$, since
the polytrope has symmetry also about the equatorial plane
$\theta=\pi/2$. The rescaled internal gravitational potential is
given by Eq. (\ref{Thetachi}), whereas the external one is given
by \beq \chi_{\rm
ext}=\nu+\sum_{l=0}^{\infty}\frac{c_l}{\xi^{l+1}}P_l(\mu)\ , \eeq
with $c_1=0,c_3=0,\ldots,c_{2k+1}=0,\ldots$. The indeterminate
coefficients $b_k$ and $c_k$ are usually evaluated by truncating
the infinite series for the potentials at a given $l$ and imposing
the matching at the surface, implicitly defined by the equation
$\Theta=0$, i.e.
\begin{eqnarray}
0&=&\int_{-1}^{1}(\chi-\chi_{\rm ext})P_{2l-2}(\mu)d\mu\ , \nonumber\\
0&=&\int_{-1}^{1}\vec n\cdot(\nabla\chi-\nabla\chi_{\rm ext})P_{2l-2}(\mu)d\mu\ ,
\end{eqnarray}
where $\vec n=\nabla\Theta$ is the normal to the surface. An
algorithm for determining the truncated set of coefficients is
presented in \cite{williams} and has been here implemented again.
For instance, for $\beta=8\times10^{-2}$ and taking terms up to
the order 8, we find: $\nu\approx-1.36398$; $b_2\approx-1.03897$;
$b_4\approx0.38613$; $b_6\approx-0.79609$; $b_8\approx-9.34534$;
$c_0\approx-4.15536$; $c_2\approx4.44831$; $c_4\approx-12.31509$;
$c_6\approx43.35993$; $c_8\approx-78.90526$. The shape of the
boundary of the star is shown in Fig. \ref{fig:surface}. We obtain
now some insights on the perturbative dynamics by studying the
geometrical properties of the associated acoustic metric.


\subsection{The effective geometry}

The acoustic metric (\ref{acousticmetric}) associated with  Williams' solution is given by the line element
\begin{eqnarray}
\label{METRIC}
ds^2&=&\frac{\rho_0}{c}[-(c^2-v_0^2)dt^2-2v_0r\sin\theta dtd\phi+dr^2+r^2(d\theta^2+\sin^2\theta d\phi^2)]\nonumber \\
&=& \frac{\rho_0}{c}[-c^2dt^2   +dr^2+r^2d\theta^2 +r^2\sin^2\theta (\Omega dt - d\phi)^2]\,,
\end{eqnarray}
where $v_0=\Omega r\sin\theta$, $c=\sqrt{2K\rho_0}$ and the
background density $\rho_0=\rho_c\Theta$. Here the coordinate
time-lines have a unit tangent vector
${1}/\sqrt{-g_{tt}}\partial_t$ ( with
$g_{tt}=-\rho_0(c^2-v_0^2)/c$) which changes its causality
relation when $v_0^2=c^2$, as it happens for instance in the case
of Minkowsky flat space-time expressed in uniformly rotating
cylindrical coordinates $(t,R,\phi,z)$, i.e. \beq
ds^2=-(C^2-\Omega^2R^2)dt^2+2\Omega R^2
dtd\phi+dR^2+R^2d\phi^2+dz^2 \label{LIGHTCYL}\eeq at the the so
called {\it light cylinder $r=C/\Omega$} where $C$ is the speed of
light\cite{LLFIELD}. Moreover in relation with metric
(\ref{METRIC}), $\partial_t$ is a Killing vector field which is
not vorticity-free (i.e. not hyper-surface orthogonal). The form
(\ref{METRIC}) of the metric suggests the introduction of the
following coordinate transformation \beq \phi' = \phi -\Omega t\
,\quad t'=t \eeq such that we obtain \beq \label{METRIC2} ds^2=
\frac{\rho_0}{c}\left(-c^2dt'^{2} +dr^2+r^2d\theta^2
+r^2\sin^2\theta d\phi'^2\right)\ . \eeq The new temporal lines
have now a unit tangent vector ${1}/\sqrt{-g_{t't'}}\partial_{t'}$
(with $g_{t't'}=-\rho_0 c$) which never changes its causality
condition inside the star. $\partial_{t'}$ is still a Killing
vector field whose norm is $-\rho_0 c$ and it is vorticity-free,
hence hyper-surface orthogonal . Such a result is not unexpected
because it is exactly what happens for the Minkowskian metric
(\ref{LIGHTCYL}) in the case of the two Killing vectors
$\partial_t$ (not vorticity free) and $\partial_t+\Omega
\partial_\phi$ (vorticity free). Since there exists everywhere a
time-like Killing vector field which is hyper-surface orthogonal,
the space-time of uniformly rotating polytropic stars under exam
is static\cite{Exact, Wald}. Clearly such a simplification due to
the uniform rotation should not apply for differentially rotating
configurations which are expected instead to lead to stationary
space-times.

Using now the dimensionless variables $\xi$ and $\eta$ defined by
\beq r=\alpha \xi\,,\qquad \eta= \sqrt{4\pi G\rho_c}\, t'\ , \eeq
with $\alpha=\sqrt{K/(2\pi G)}$, leads to the following final form
of the metric
\begin{eqnarray}
\label{METRIC3}
ds^2
&=&\sqrt{\Theta }\left[-\Theta d\eta^2 +d\xi^2+\xi^2(d\theta^2 +\sin^2\theta  d\phi^2)\right]\ ,
\end{eqnarray}
where we have replaced $\phi'$ by $\phi$ for convenience and the
ignorable constant multiplicative factor $2\pi G\sqrt{K\rho_c/2}$
has been dropped, since it can be re-absorbed in the definitions
of $\eta$ and $\xi$ by a simple rescaling of such variables. The
metric determinant is given by $g=-\Theta^3\xi^4\sin^2\theta$,  so
that $\sqrt{-g}=\Theta^{3/2}\xi^2\sin\theta$, which implies that
the volume element vanishes approaching both the center of the
configuration where $\Theta=1$ and the boundary where $\Theta=0$,
and it is well behaved otherwise.

\begin{figure}
\typeout{*** EPS figure kretch}
\begin{center}
\includegraphics[scale=0.5]{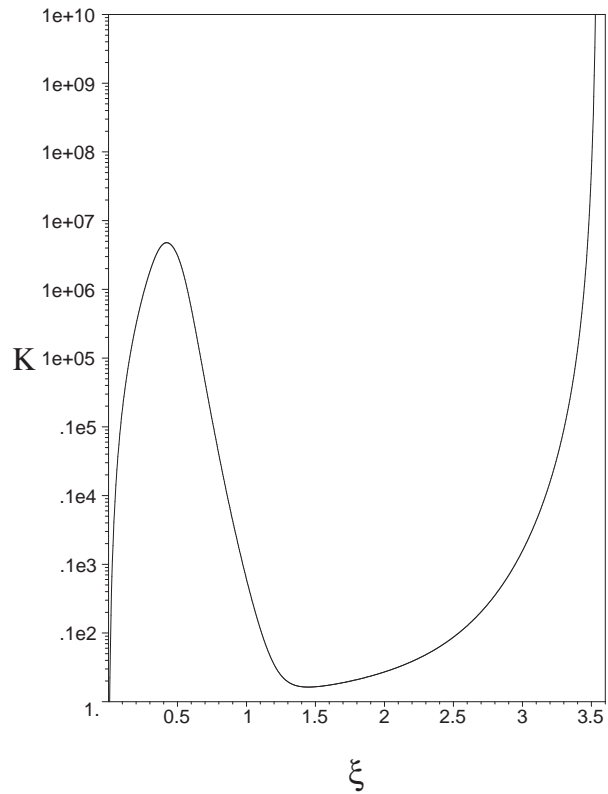}
\end{center}
\caption{The behavior of the Kretschmann invariant ${\cal K}$ is shown as a function of $\xi$ for $\beta=8\times10^{-2}$ and fixed $\theta=\pi/3$.
The boundary is located at $\xi\approx3.54$.
${\cal K}$ diverges there, whereas gets a constant value at the center.
}
\label{fig:kretch}
\end{figure}

It is useful to evaluate the Kretschmann invariant\cite{Exact}
${\cal K}=R_{\alpha\beta\gamma\delta}R^{\alpha\beta\gamma\delta}$.
It diverges at the boundary, whereas gets a constant value at the
center (see Fig. \ref{fig:kretch}). In fact, its behavior for
$\xi\to0$ is given by \beq {\cal
K}\simeq\frac{2b_2^2}{15\pi}+\frac{13}{12}(1-\beta)^2+O(\xi^2)\,.
\eeq Other relevant curvature invariants are listed in Appendix
\ref{app1}.


It is worth noting that curvature invariants give useful informations about the
presence of possible curvature pathologies, whereas the analysis of
the metric tensor and its determinant (the latter
being a tensor density) do not contain intrinsic information
about the spacetime, but only coordinate-dependent informations.

The equations governing geodesic motion are given by
\begin{subequations}
\begin{eqnarray}
\label{eqeta}
\dot \eta&=&\frac{E}{\Theta^{3/2}}\,, \\
\label{eqphi}
\dot \phi&=&\frac{L}{\sqrt{\Theta}\xi^2\sin^2\theta}\,, \\
\label{eqxi}
{\dot \xi}^2&=&-\xi^2{\dot \theta}^2-\frac{\sigma^2}{\sqrt{\Theta}}-\frac{L^2}{\xi^2\Theta\sin^2\theta}+\frac{E^2}{\Theta^{2}}\,, \\
\label{eqtheta}
\ddot \theta&=&-\frac12\left(\frac{4}{\xi}+\frac{\partial_\xi\Theta}{\Theta}\right){\dot \xi}{\dot \theta}-\frac12\left[{\dot \theta}^2+\frac{\sigma^2}{2\xi^2\sqrt{\Theta}}+\frac{E^2}{\xi^2\Theta^{2}}\right]\frac{\partial_\theta\Theta}{\Theta}+\frac{\cos\theta}{\sin^3\theta}\frac{L^2}{\xi^4\Theta}\,,
\end{eqnarray}
\end{subequations}
where Killing symmetries and the normalization condition have been
used. Here $E$ and $L$ are constants of motion (representing
energy and angular momentum), $\sigma^2=1,0,-1$ for timelike,
null and spatial geodesics respectively and a dot denotes
differentiation with respect to the affine parameter.

Let us consider first the motion on the equatorial plane
$\theta=\pi/2$. If $\theta=\pi/2$ and $\dot \theta=0$ initially,
Eq. (\ref{eqtheta}) ensures that the motion will be confined
on the equatorial plane, since $\partial_\theta\Theta$ vanishes at
$\theta=\pi/2$, so that $\ddot \theta=0$ too.
The geodesic equations thus reduce to
\begin{eqnarray}
\label{geoeqnsequat}
\dot \eta&=&\frac{E}{\Theta^{3/2}}\,, \qquad
\dot \phi=\frac{L}{\sqrt{\Theta}\xi^2}\,, \qquad
{\dot \xi}^2=-\frac{\sigma^2}{\sqrt{\Theta}}-\frac{L^2}{\xi^2\Theta}+\frac{E^2}{\Theta^{2}}\,,
\end{eqnarray}
where the function $\Theta$ is meant to be evaluated at
$\theta=\pi/2$. The $\xi$--motion turns out to be
governed by the effective potential $V_{\xi}$, implicitly defined
by \beq \label{eqVxi}
-\frac{\sigma^2}{\sqrt{\Theta}}-\frac{L^2}{\xi^2\Theta}+\frac{V_{\xi}^2}{\Theta^{2}}=0\,.
\eeq
In fact, for $E=V_{\xi}$ the right hand side of the last equation of Eq.
(\ref{geoeqnsequat}) vanishes.

Similarly, one can  introduce the effective potential governing the
$\theta$-motion, i.e. with
$\xi=\,$const:
\beq
\label{eqVtheta}
-\frac{\sigma^2}{\sqrt{\Theta}}-\frac{L^2}{\xi^2\Theta\sin^2\theta}+\frac{V_{\theta}^2}{\Theta^{2}}=0\,,
\eeq
as from Eq. (\ref{eqxi}) where we set $\dot\xi=0$.

\subsection{Perturbations}

The set of equations governing fluctuations about the unperturbed
reference flow are given by Eq. (\ref{FINALSRISCR}). This is a
system of coupled PDEs which cannot be solved analytically. Even a
direct numerical integration in the time domain is a hard task as
discussed later.

The problem can be simplified by noting that the contribution of $\vec\xi_1$ to the first order velocity
field is generally a small correction with respect to $\nabla\psi_1$, as discussed in Ref. \cite{ROTA}.
The system (\ref{FINALSRISCR}) thus becomes
\begin{eqnarray}
\nabla^\mu\nabla_\mu\psi_1&=&
-\frac{c}{\rho_0}\left(\frac{D^{(0)}}{d
t}\right)\left(\frac{\Phi_1}{c^2}\right)
\,,\nonumber\\
 \left[\nabla^2+k_J^2\right] \Phi_1&=&-k_J^2
\frac{D^{(0)}}{dt}\psi_1\,,
\end{eqnarray}
with $\vec\xi_1$ neglected.
It is worth to stress that the quantity $\vec\xi_1$
represents the correction to potential flow induced by angular
momentum conservation, with a partial hybridization of the sound
with other modes. At low frequencies (see Ref. \cite{ROTA}) the contribution by $\vec\xi_1$ ceases to be negligible
and the sound waves hybridize with the many other modes
available for a fluid whose vorticity can have comparable
frequency. In our analysis, however, we shall be far from this scenario.

 By neglecting also the
gravitational back-reaction, linear perturbations of the velocity
potential satisfy the Klein-Gordon equation for a massless scalar
field $\nabla^\mu\nabla_\mu\psi_1=0$ on the background metric
(\ref{METRIC3}), i.e. \beq \label{eqpsi1}
\left[-\frac{\partial^2}{\partial
\eta^2}+\Theta\nabla^2+\partial_\xi\Theta\frac{\partial}{\partial\xi}+\frac{\partial_\theta\Theta}{\xi^2}\frac{\partial}{\partial\theta}\right]\psi_1=0\
, \eeq where $\nabla^2$ is the ordinary Laplacian in spherical
coordinates. Unfortunately this equation is not completely
separable, so it should be studied numerically.

\begin{figure}
\typeout{*** EPS figure rad1}
\begin{center}
$\begin{array}{cc}
\includegraphics[scale=0.5]{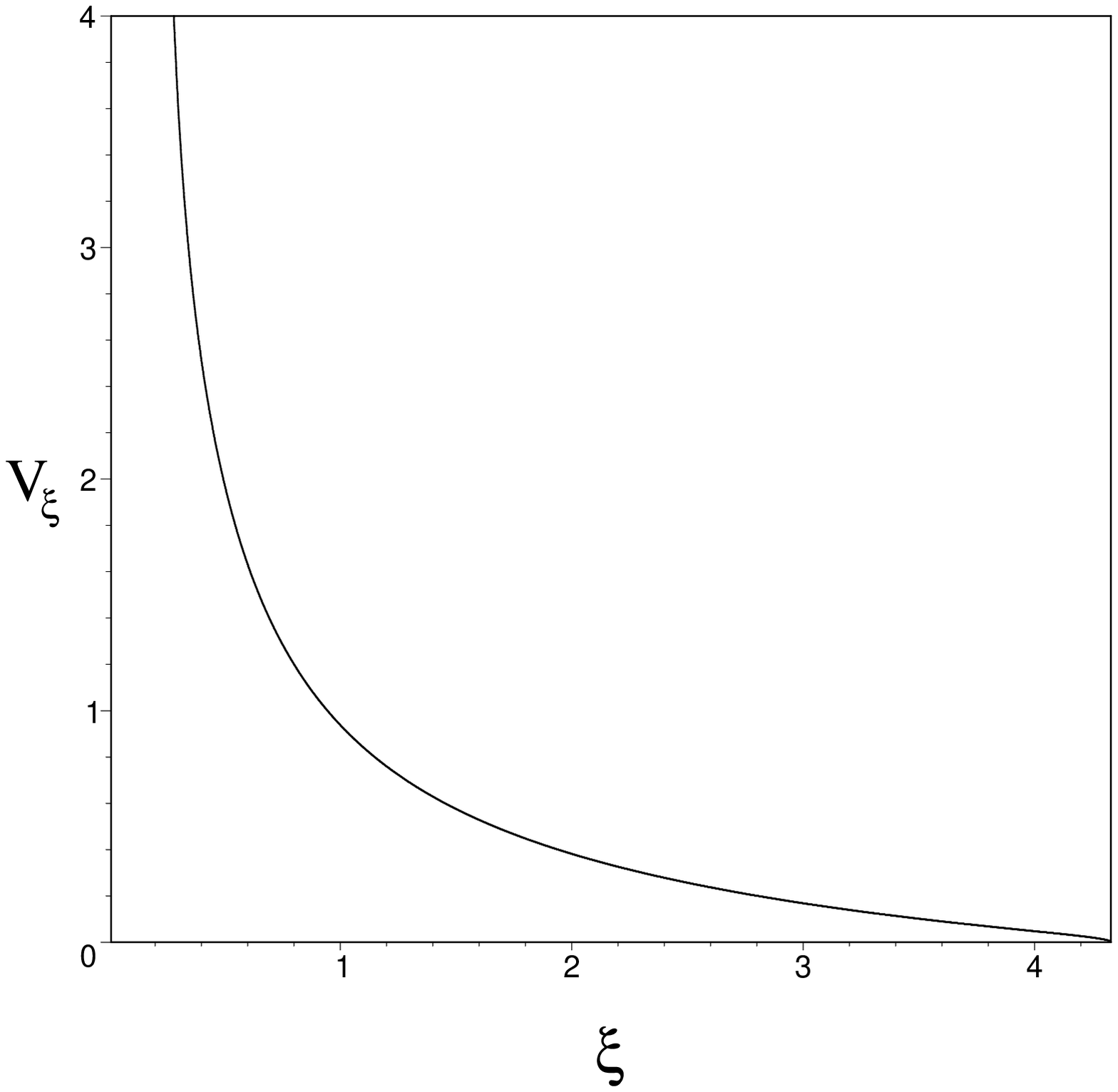}&
\includegraphics[scale=0.5]{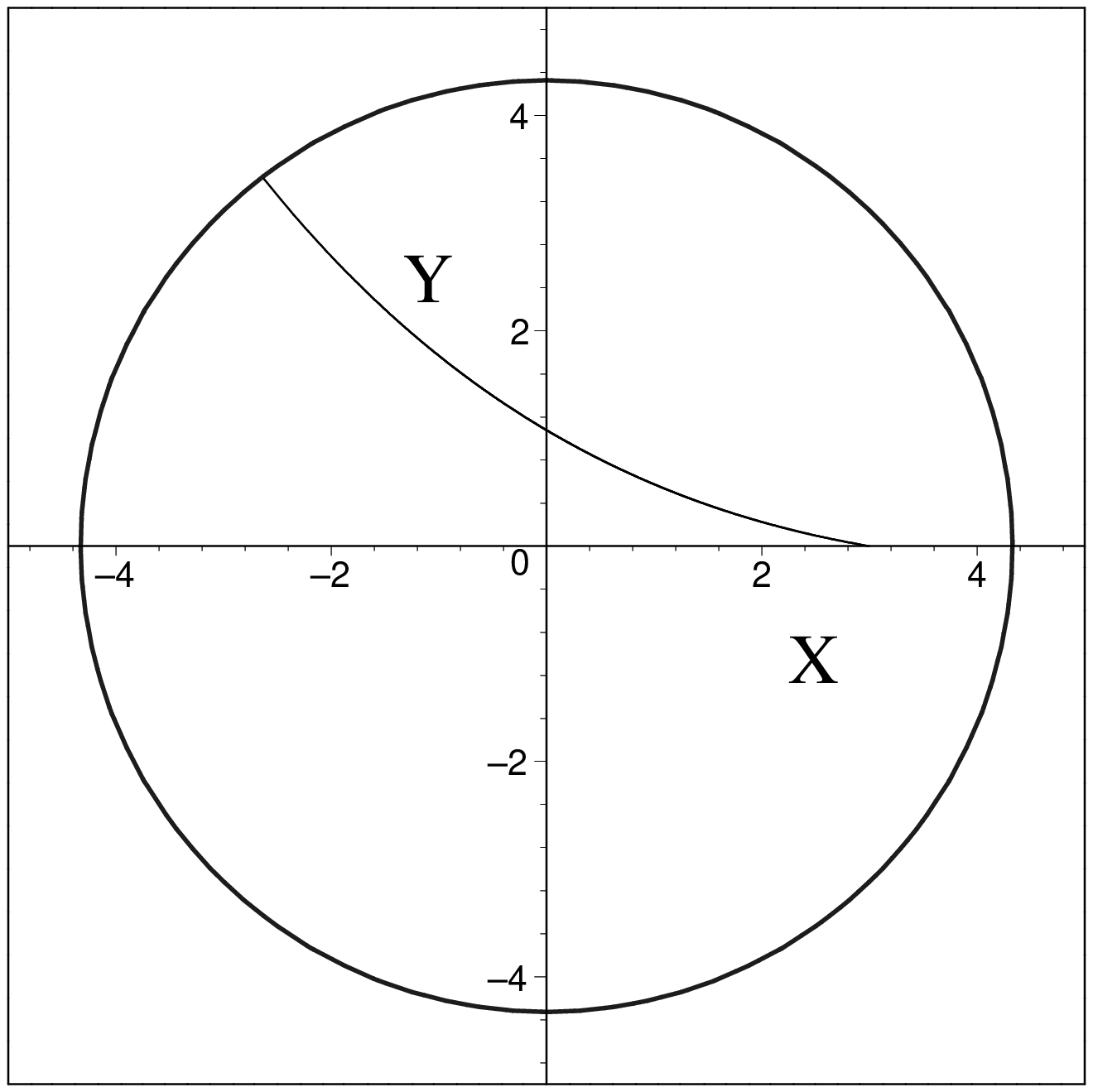}\\[.4cm]
\quad\mbox{(a)}& \quad\mbox{(b)}
\end{array}$\\
\end{center}
\caption{The effective potential $V_\xi$ for radial motion of null rays is shown in Fig. (a) for $\theta=\pi/2$ and fixed values of $\beta=8\times10^{-2}$ and $L=1$.
Fig. (b) shows the path of a sound ray starting from the equatorial plane with energy $E=1$ and angular momentum $L=1$ at $X=3$.
}
\label{fig:rad1}
\end{figure}

\begin{figure}
\typeout{*** EPS figure spher1}
\begin{center}
$\begin{array}{cc}
\includegraphics[scale=0.45]{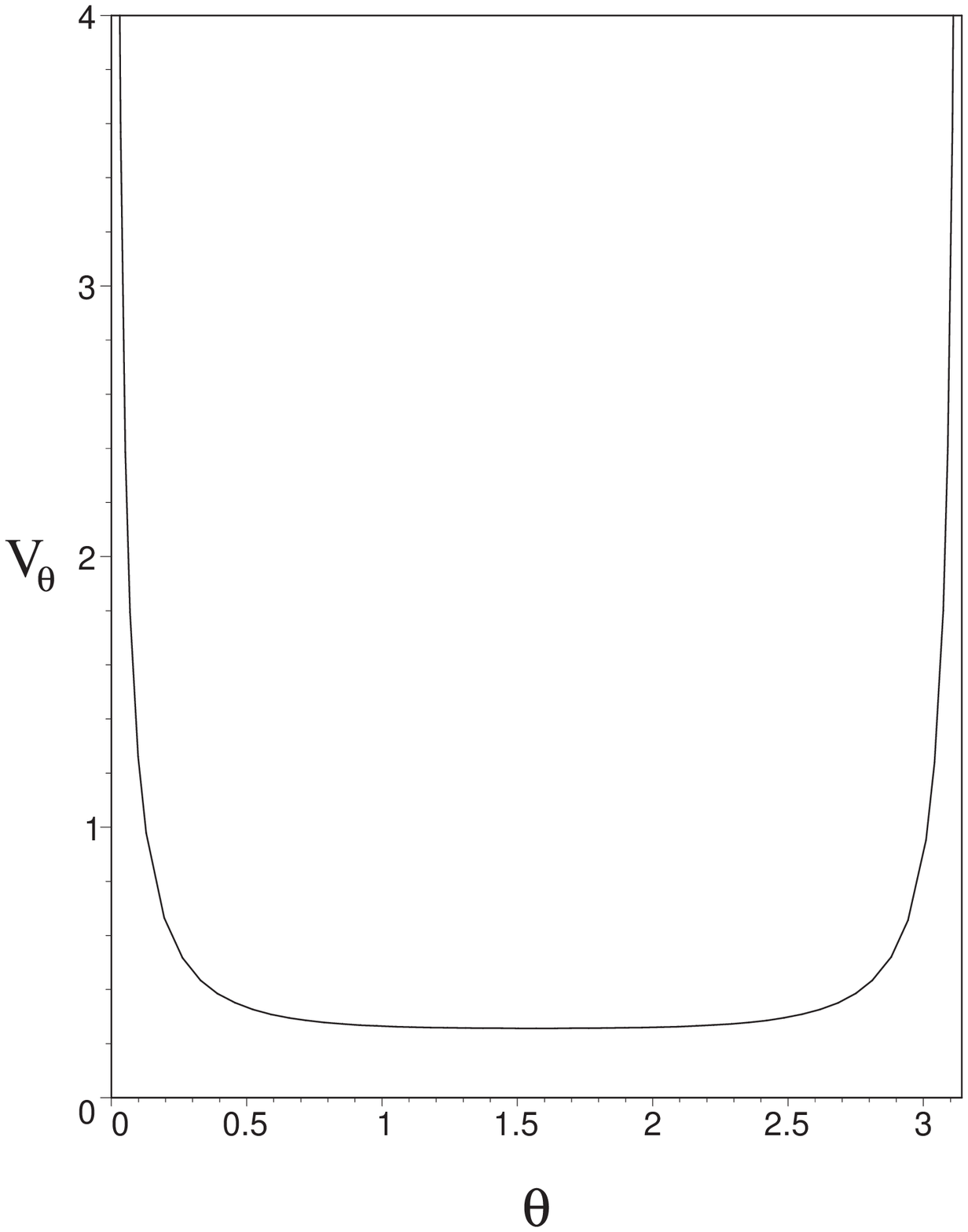}&\quad
\includegraphics[scale=0.5]{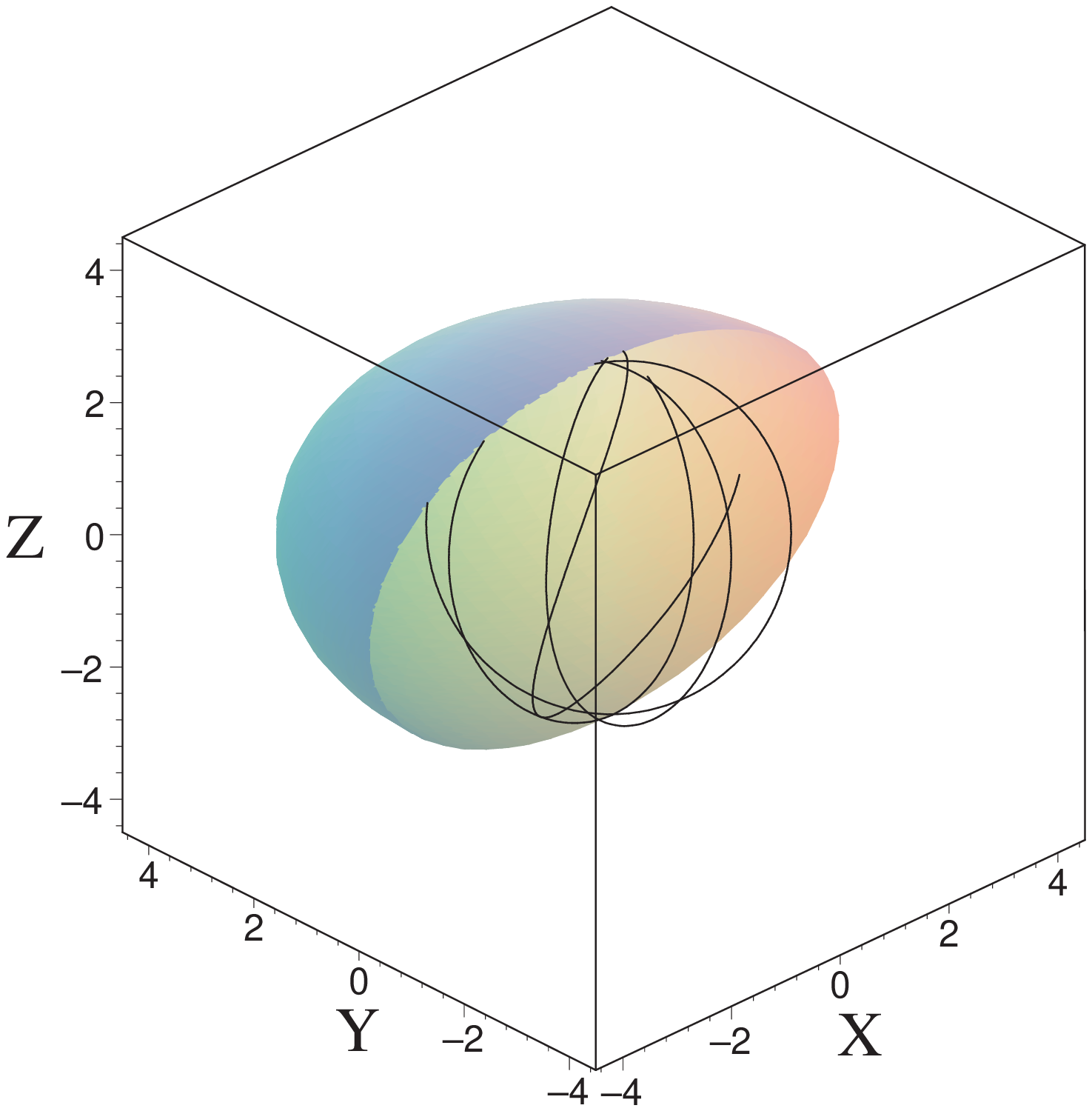}\\[.4cm]
\quad\mbox{(a)}\quad &\quad \mbox{(b)}\\[.6cm]
\includegraphics[scale=0.45]{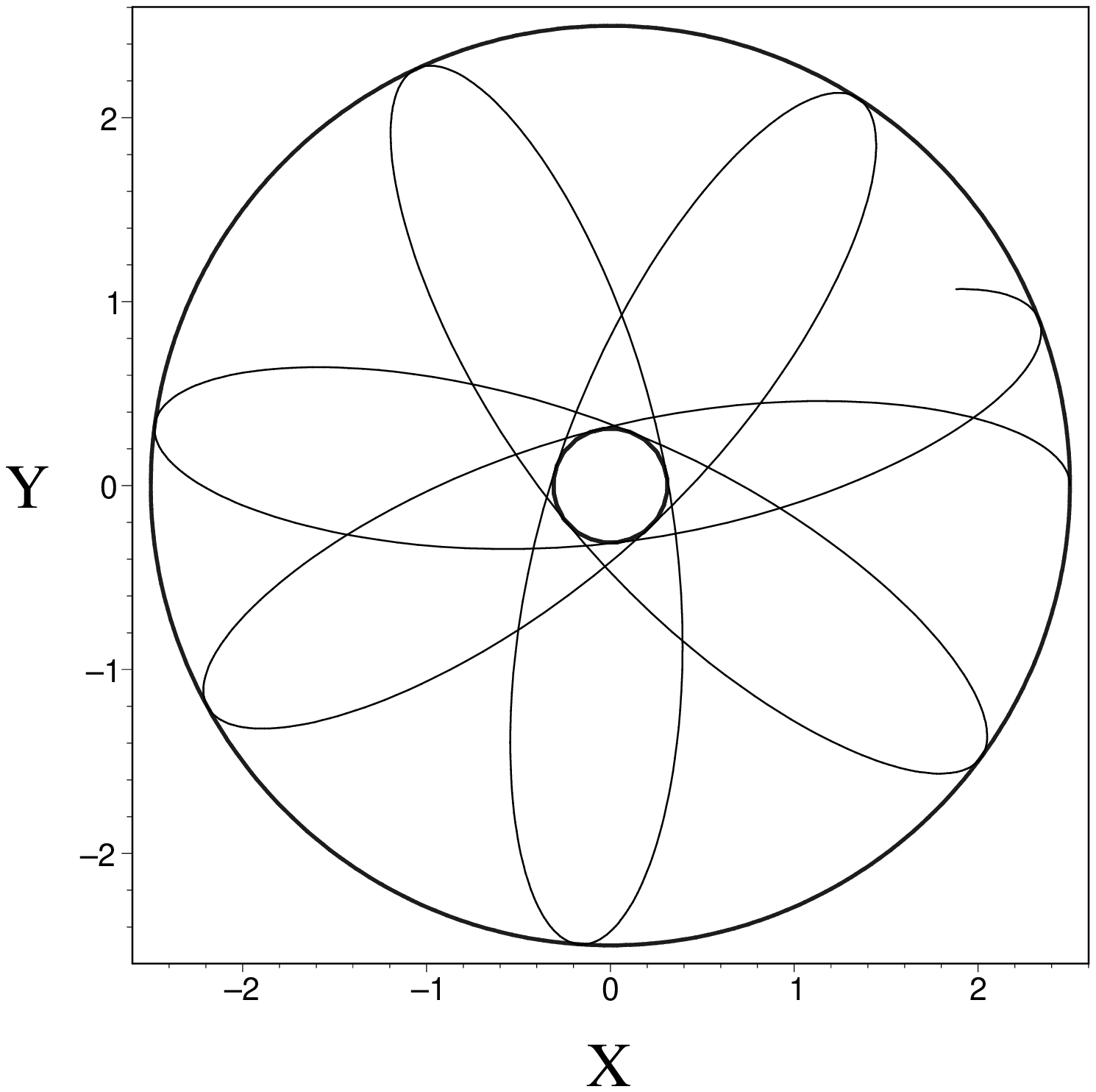}&\quad
\includegraphics[scale=0.45]{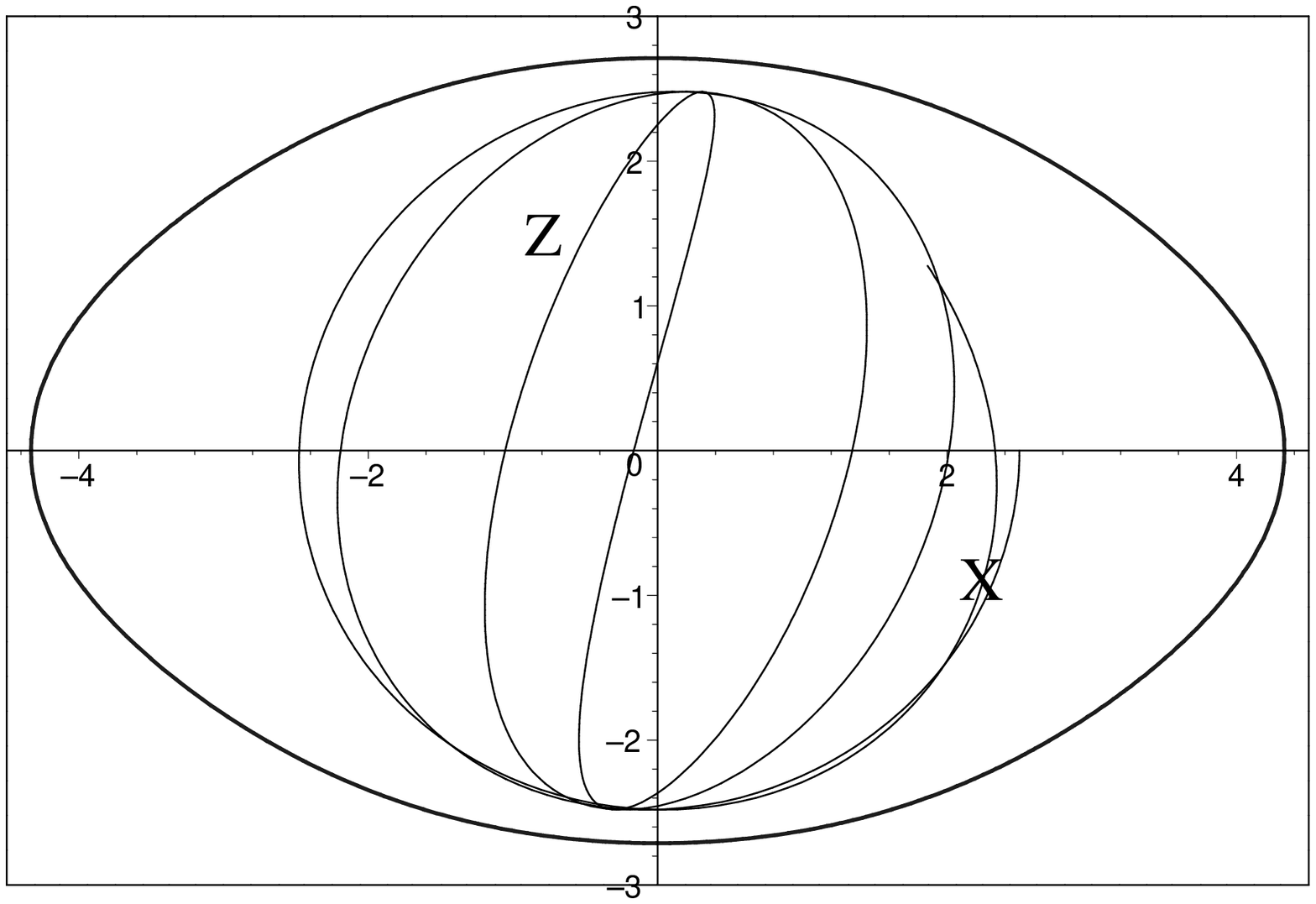}\\[.4cm]
\quad\mbox{(c)}\quad &\quad \mbox{(d)}
\end{array}$\\
\end{center}
\caption{The effective potential $V_\theta$ for polar motion of null rays is shown in Fig. (a) for $\xi=2.5<\xi_-$ and fixed values of $\beta=8\times10^{-2}$ and $L=1$ ($X=\xi\sin\theta\cos\phi$, $Y=\xi\sin\theta\sin\phi$ and $Z=\xi\cos \theta$ are Cartesian-like coordinates).
The boundary surface has been cut in half for a better view of the interior.
The corresponding spherical orbit of a ray starting from the equatorial plane with energy $E=1$ is shown in Fig. (b).
The projected path on both planes $X-Y$ and $X-Z$ is shown in Figs. (c) and (d), respectively.
}
\label{fig:spher1}
\end{figure}

\begin{figure}
\typeout{*** EPS figure spher2}
\begin{center}
$\begin{array}{cc}
\includegraphics[scale=0.45]{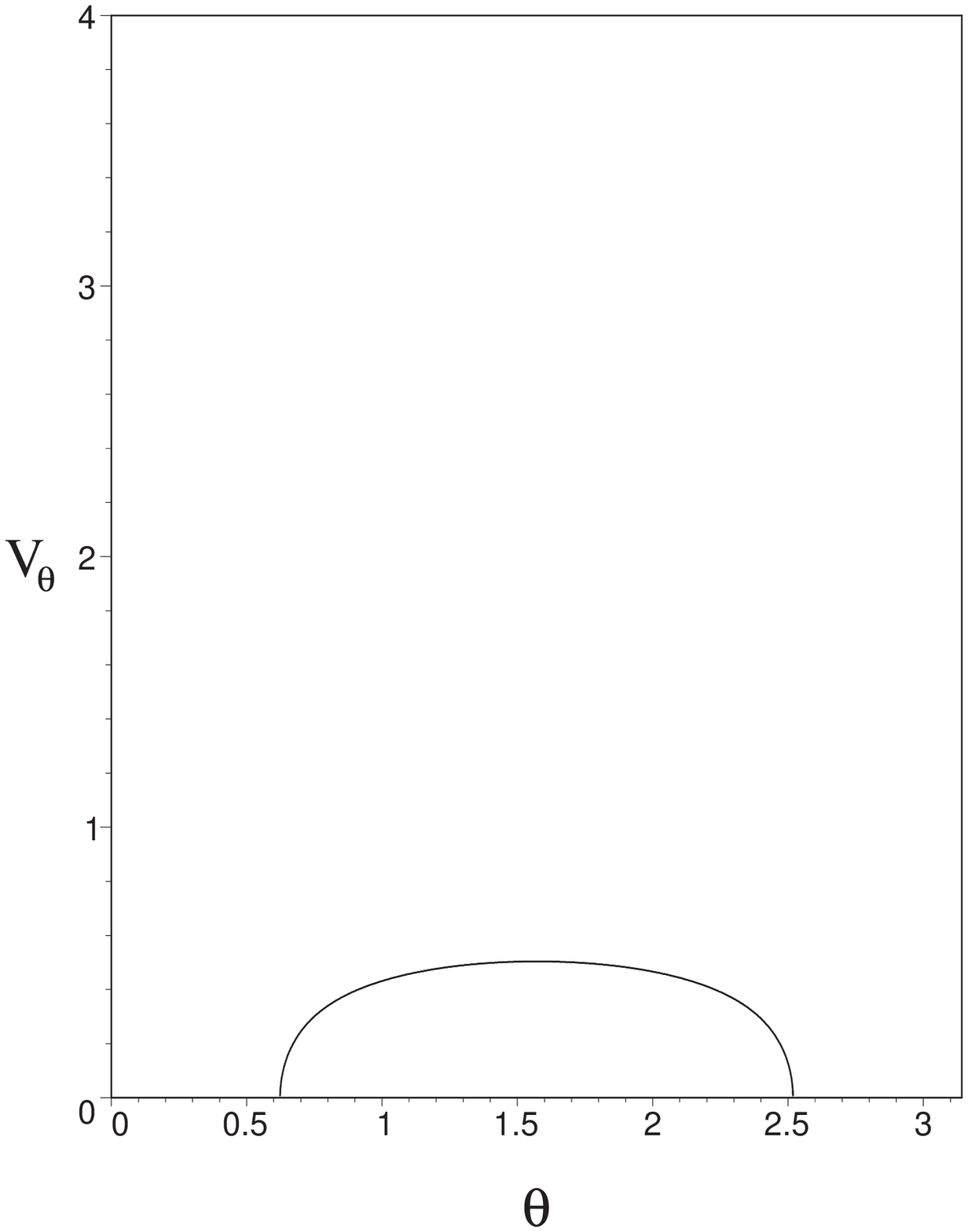}&\quad
\includegraphics[scale=0.5]{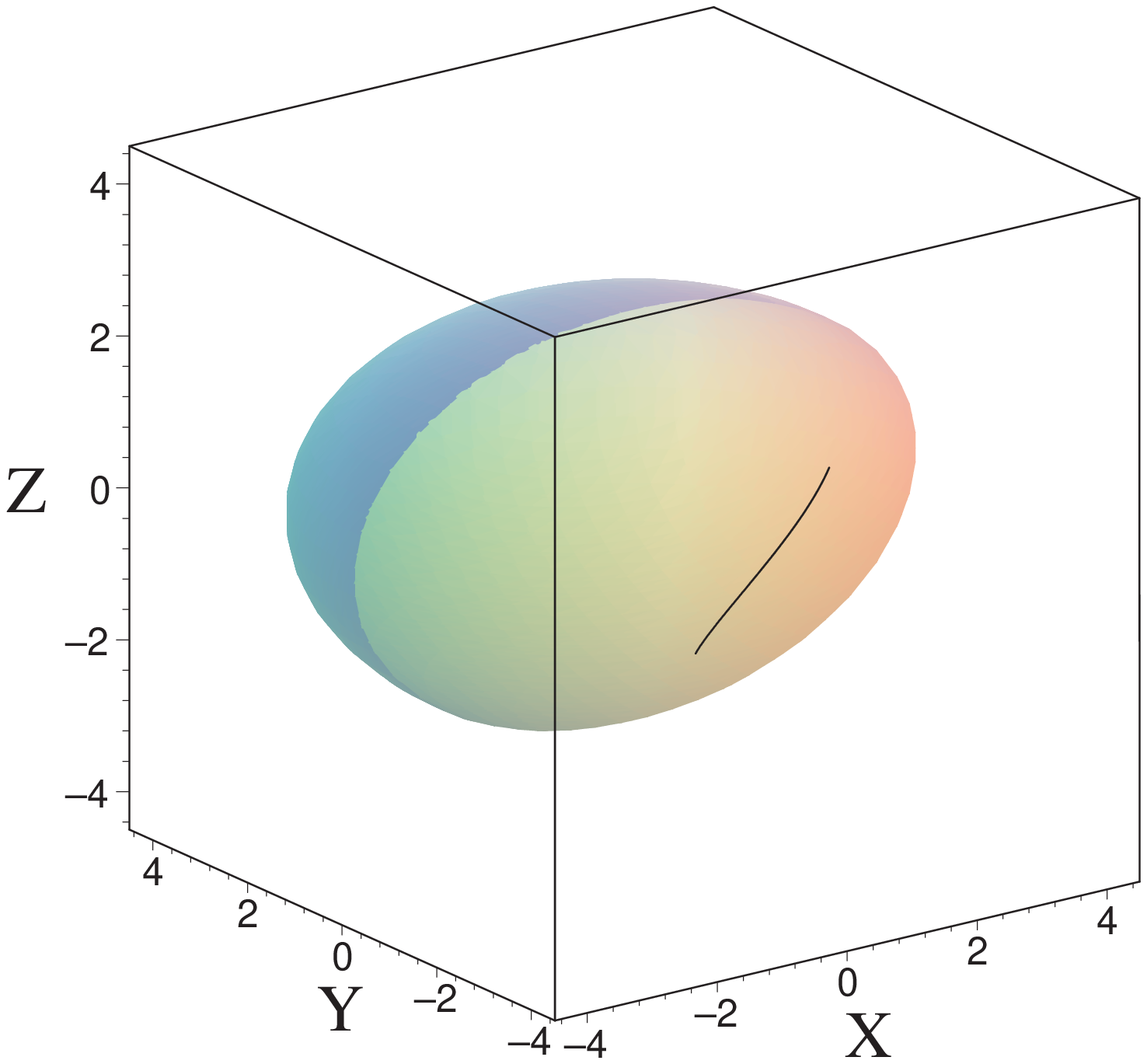}\\[.4cm]
\quad\mbox{(a)}\quad &\quad \mbox{(b)}\\[.6cm]
\includegraphics[scale=0.45]{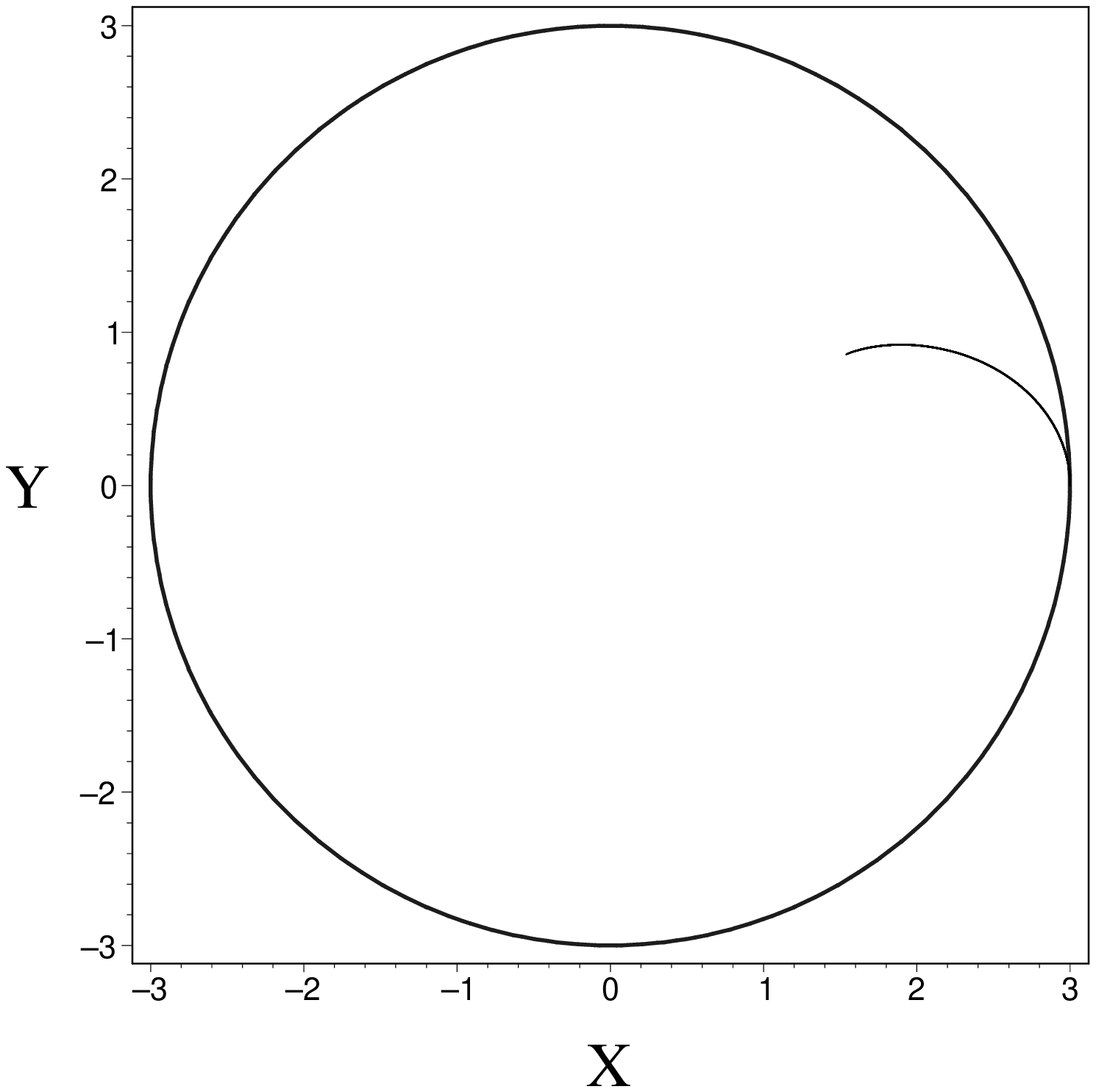}&\quad
\includegraphics[scale=0.45]{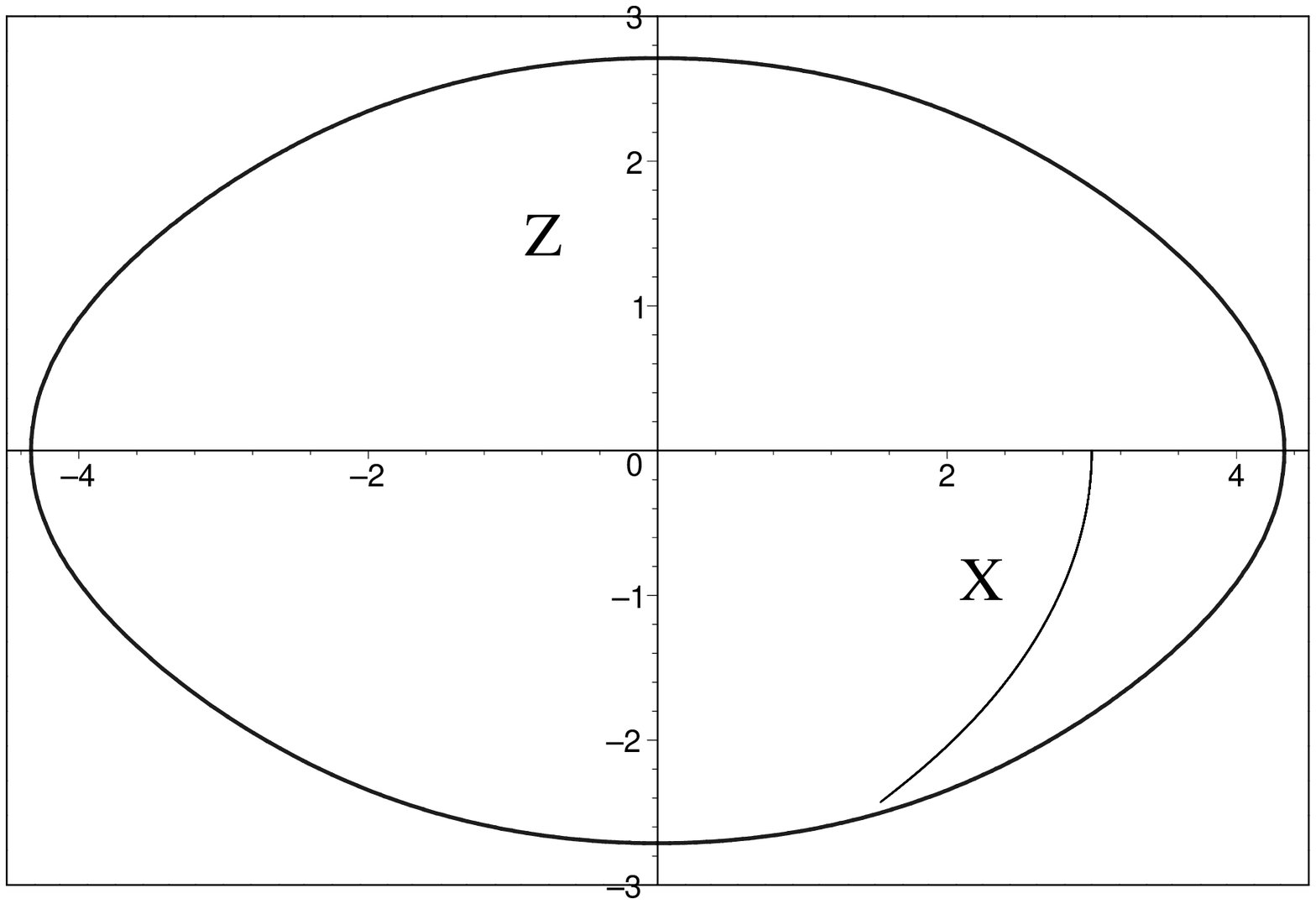}\\[.4cm]
\quad\mbox{(c)}\quad &\quad \mbox{(d)}
\end{array}$\\
\end{center}
\caption{The effective potential $V_\theta$ for polar motion of null rays is shown in Fig. (a) for $\xi=3>\xi_-$ and fixed values of $\beta=8\times10^{-2}$ and $L=3$.
The corresponding spherical orbit of a ray starting from the equatorial plane with energy $E=1$ is shown in Fig. (b).
The projected path on both planes $X-Y$ and $X-Z$ is shown in Figs. (c) and (d), respectively.
The numerical integration stops when the boundary of the configuration is reached.
}
\label{fig:spher2}
\end{figure}

However, one expects that the analog light cone (or sound cone) structure of our problem resulting from the numerical integration of the wave equation (\ref{eqpsi1}) is related to null geodesics (i.e. Eqs.
(\ref{eqeta})--(\ref{eqtheta}) with $\sigma=0$) in a standard way, in the sense
that these are the high frequency limit of $\psi_1$.
Fig. \ref{fig:rad1} (a) shows the behavior of the effective potential $V_\xi$ for radial motion of null rays on the equatorial plane $\theta=\pi/2$ and fixed values of the rotation parameter $\beta$.
We see that the center of the configuration is approached only by high energy rays.
Fig. (b) shows instead the typical path of a sound ray starting from the equatorial plane at a given distance from the center with fixed energy and angular momentum.
The behavior of the effective potential $V_\theta$ for polar motion of null rays is shown in Figs. \ref{fig:spher1} (a) and \ref{fig:spher2} (a) for fixed values of the radial parameter $\xi<\xi_-$ and $\xi>\xi_-$, respectively.
In the first case the motion is confined between two values of the polar angle for fixed energy, whereas in the latter case the ray can reach the boundary of the configuration.
The corresponding spherical orbits of a ray starting from the equatorial plane with fixed energy and angular momentum are shown in Figs. \ref{fig:spher1} (b)--(d) and \ref{fig:spher2} (b)--(d), respectively.

\begin{figure}
\typeout{*** EPS figure cone}
\begin{center}
\includegraphics[scale=0.5]{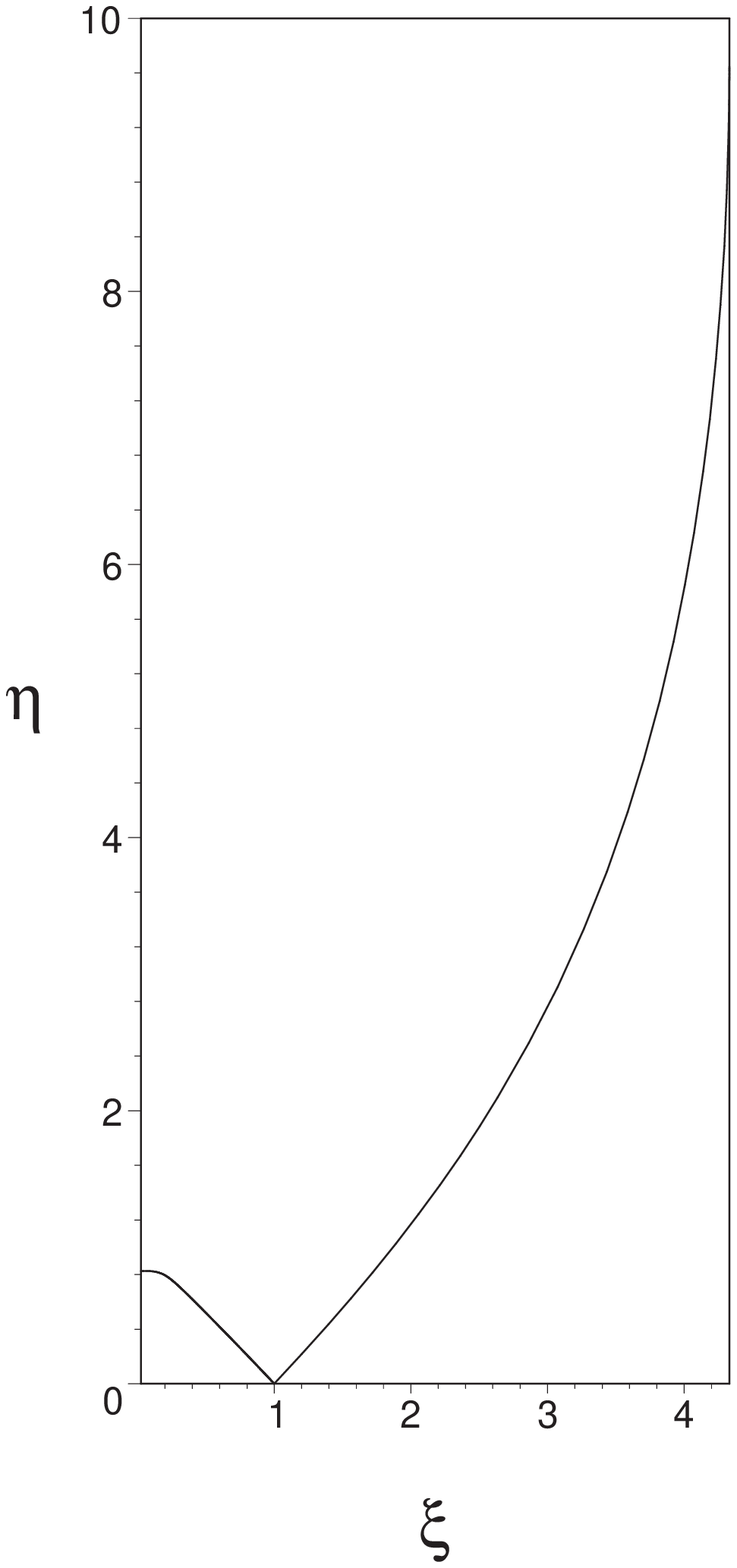}
\end{center}
\caption{Analog light cone structure for radially accelerated sound rays starting at the point $\xi=1$ for $\eta=0$ from the equatorial plane $\theta=\pi/2$ for $\beta=8\times10^{-2}$.
The boundary $\xi\approx4.33$ is reached at a finite time.
}
\label{fig:cone}
\end{figure}

The analog light cone structure is obtained by setting $ds^2=0$ in
Eq. (\ref{METRIC3}). For instance, the equation governing the
motion of accelerated null rays moving radially on the equatorial
plane turns out to be given by \beq \label{eqcone}
\frac{d\xi}{d\eta}=\pm\sqrt{\Theta}\ , \eeq where $\Theta$ is
evaluated at $\theta=\pi/2$. The sound cone structure arising
from the numerical integration of Eq. (\ref{eqcone}) is shown in
Fig. \ref{fig:cone}. Note that the boundary is reached at a finite
time as in the spherical case \cite{bcfprd}.


\section{Concluding remarks}

In this paper we have derived the field equations for general
acoustic perturbations of a perfect rotating and self-gravitating
compressible gaseous/fluid mass through an extended analog
model based on Clebsch potentials, here generalized to
account  for gravitational backreaction.

We have then examined in detail the case of the uniformly rotating
$n=1$ polytropic configuration as an analytical background  mainly
focusing on the geometric properties of the associated acoustic
metric.

The surface of the star, as in the spherical
case \cite{bcfprd}, corresponds to a zero density
condition, so that the sound speed vanishes, while the curvature of
the spacetime diverges (the geometry breaks down manifesting a
curvature singularity). A simple explanation of the curvature
divergence, even in presence of rotation, is that acoustics
require a medium on which waves travel. The fluid is the source of
the induced metric tensor, but on the surface of the star there is
no medium, so there cannot be sound too: geometric analogies fail here.

The presence of rotation leads in general
to a stationary problem with an ergosphere where the fluid
velocity exceeds the local speed of sound. Stationarity, however,
must be associated with a differential rotation of the fluid. In
the simplest uniformly rotating case considered here the
metric is indeed static, leading to a relatively simple dynamics.
On the other hand the case of differential rotation should provide
more complicated situations in which such a simplification could
not be achievable anymore.

Our study could be useful to better understand the non
relativistic theory of stellar and galactic structures via methods
typical of General Relativity. In particular, a problem which
still remains to be studied is how, for uniformly or
differentially rotating polytropes, our perturbative geometrized
formulation can describe instabilities leading to bifurcations
similar to those already discussed in the literature using
different mathematical approaches (see e.g. Ref. \cite{Tholine}).
In this context in particular it would be important to see if in
case of outer fluid layers rotating with supersonic velocities,
there could be a superradiant scattering of sound waves, already
evidenced in acoustic black holes (see Ref. \cite{CHR1} and
references therein) as it happens for astrophysical systems as
black holes. This possible analysis requires however an important
technical clarification. In black hole physics in fact the
perturbation analysis in search of instabilities can be performed
by analytical techniques due to the complete separability of the
perturbation equation in radial and angular variables in the
frequency domain, leading to a Schroedinger-like problem
\cite{Chandra}. Unfortunately, in the present context such a
simplification does not occur due to the non variables
separability of density and speed of sound so the whole analysis
then can only be performed numerically including the ordinary
differential equations of geodesics discussed in this article. The
perturbative equations in particular are partial differential
equations leading to an involved study which would require in the
time domain the implementation of specific numerical codes in
$\lq\lq 3+1"$ or $\lq\lq 2+1"$ (using axisymmetry of the
background) dimensions written using modern numerical relativity
tools, and therefore is left to future works on the lines of Refs.
\cite{CHR1,CHR2}.

\begin{acknowledgments}
The authors acknowledge ICRANet for support.
\end{acknowledgments}

\appendix

\section{Newman-Penrose quantities}
\label{app1}

The following  Newman-Penrose frame (we follow Ref. \cite{Frolov} for conventions  here) allows one to study in detail
the curvature structure of the acoustic metric (\ref{METRIC3}):
\beq
l= \frac{1}{\sqrt{2}}(e_{\hat \eta}+e_{\hat \xi})\,,\qquad
n=\frac{1}{\sqrt{2}}(e_{\hat \eta}-e_{\hat \xi})\,,\qquad
m=\frac{1}{\sqrt{2}}(e_{\hat \theta}+ie_{\hat \phi})\ ,
\eeq
where
\beq
e_{\hat \eta}=\frac{1}{\sqrt{-g_{\eta\eta}}}\partial_\eta\ , \quad
e_{\hat \xi}=\frac{1}{\sqrt{g_{\xi\xi}}}\partial_\xi\ , \quad
e_{\hat \theta}=\frac{1}{\sqrt{g_{\theta\theta}}}\partial_\theta\ , \quad
e_{\hat \phi}=\frac{1}{\sqrt{g_{\phi\phi}}}\partial_\phi\ .
\eeq
The nonvanishing spin coefficients are
\begin{eqnarray}
\kappa&=&\frac{1}{4\sqrt{2}}\frac{\Theta_\theta}{\Theta^{5/4}\xi}=\frac{\tau}{2}=\nu=\frac{\pi}{2}\
, \nonumber\\
\epsilon&=&-\frac{3}{8\sqrt{2}}\frac{\Theta_\xi}{\Theta^{5/4}}=\gamma\
, \nonumber\\
\alpha&=&\frac{\kappa}{2}+\frac{1}{2\sqrt{2}}\frac{\cot\theta}{\Theta^{1/4}\xi}=-\beta\
, \nonumber\\
\rho&=&-\frac23\epsilon+\frac{1}{\sqrt{2}}\frac{1}{\Theta^{1/4}\xi}=-\mu\
.
\end{eqnarray}
The Weyl scalar in this frame are
\begin{eqnarray}
\psi_0&=&\psi_4=\frac{1}{8\Theta^{3/2}\xi^2}\left[\Theta_{\theta\theta}-\Theta_\theta\left(\frac{\Theta_\theta}{2\Theta}+\cot\theta\right)\right]\
, \nonumber\\
\psi_1&=&-\psi_3=-\frac{1}{8\Theta^{3/2}\xi}\left[\Theta_{\xi\theta}-\Theta_\theta\left(\frac{\Theta_\xi}{2\Theta}+\frac{1}{\xi}\right)\right]\
, \nonumber\\
\psi_2&=&\frac{1}{6\Theta^{3/2}}\left[\Theta_{\xi\xi}-\frac{\Theta_{\theta\theta}}{2\xi^2}-\Theta_\xi\left(\frac{\Theta_\xi}{2\Theta}+\frac{1}{\xi}\right)+\frac{\Theta_{\theta}}{2\xi^2}\left(\frac{\Theta_\theta}{2\Theta}-\cot\theta\right)\right]\
.
\end{eqnarray}
Other nonvanishing NP quantities are the curvature scalar
\beq
R=\frac{5}{2\Theta^{3/2}}\left[\Theta_{\xi\xi}+\frac{\Theta_{\theta\theta}}{\xi^2}-2\Theta_\xi\left(\frac{7}{40}\frac{\Theta_\xi}{\Theta}-\frac{1}{\xi}\right)-\frac{\Theta_{\theta}}{\xi^2}\left(\frac{7}{20}\frac{\Theta_\theta}{\Theta}-\cot\theta\right)\right]\,,
\eeq
and the Ricci coefficients
\begin{eqnarray}
\Phi_{00}&=&\Phi_{22}=\frac{1}{8\Theta^{3/2}}\left[\Theta_{\xi\xi}-\frac{\Theta_{\theta\theta}}{\xi^2}-2\Theta_\xi\left(\frac{7}{8}\frac{\Theta_\xi}{\Theta}+\frac{1}{\xi}\right)-\frac{\Theta_{\theta}}{\xi^2}\cot\theta\right]\
, \nonumber\\
\Phi_{11}&=&-\Phi_{00}-\frac{R}{40}-\frac{21}{160\Theta^{5/2}}\left(\Theta_\xi^2+\frac{\Theta_\theta^2}{\xi^2}\right)\
, \nonumber\\
\Phi_{01}&=&-\Phi_{12}=\frac{1}{4\Theta^{3/2}\xi}\left[\Theta_{\xi\theta}-\Theta_\theta\left(\frac{7}{8}\frac{\Theta_\xi}{\Theta}+\frac{1}{\xi}\right)\right]\
, \nonumber\\
\Phi_{02}&=&\frac{1}{4\Theta^{3/2}\xi^2}\left[\Theta_{\theta\theta}-\Theta_{\theta}\left(\frac{7}{8}\frac{\Theta_\theta}{\Theta}+\cot\theta\right)\right]\
.
\end{eqnarray}

\begin{figure}
\typeout{*** EPS figure weylscal}
\begin{center}
\includegraphics[scale=0.4]{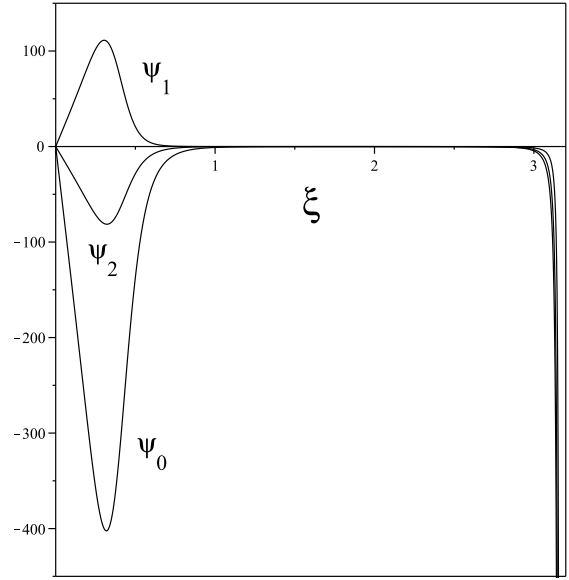}
\end{center}
\caption{The independent Weyl scalars $\psi_0$, $\psi_1$ and $\psi_2$ are plotted as functions of $\xi$ for $\beta=8\times10^{-2}$ and fixed $\theta=\pi/4$.
The boundary is located at $\xi\approx3.17$.
They all diverge there, whereas get a constant value at the center.
}
\label{fig:weylscal}
\end{figure}

\begin{figure}
\typeout{*** EPS figure ricciscal}
\begin{center}
\includegraphics[scale=0.4]{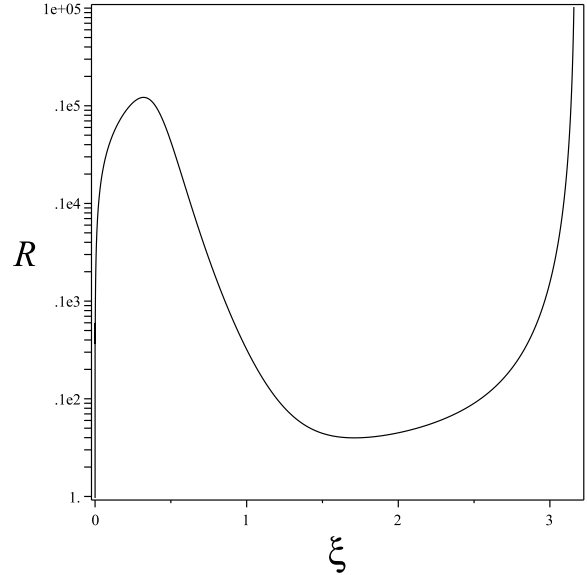}
\end{center}
\caption{The behavior of the Ricci scalar $R$ is shown as a function of $\xi$ for $\beta=8\times10^{-2}$ and fixed $\theta=\pi/4$.
It diverges at the boundary ($\xi\approx3.17$), whereas gets a constant value at the center.
}
\label{fig:ricciscal}
\end{figure}

The behavior of the independent Weyl scalars as well as the Ricci scalar is shown as a function of $\xi$ for fixed $\theta$ in Figs. \ref{fig:weylscal} and \ref{fig:ricciscal}, respectively.
All quantites diverge at the boundary, whereas get a constant value at the center. In fact, for $\xi\to0$ we have
\begin{eqnarray}
\psi_0&\simeq&\frac1{20}\frac{b_2}{\sqrt{2\pi}}\sin^2\theta+O(\xi^2)\,, \nonumber\\
\psi_1&\simeq&\frac1{20}\frac{b_2}{\sqrt{2\pi}}\sin\theta\cos\theta+O(\xi^2)\,, \nonumber\\
\psi_2&\simeq&\frac1{60}\frac{b_2}{\sqrt{2\pi}}(3\cos^2\theta-1)+O(\xi^2)\,, \nonumber\\
R&\simeq&\frac{5}{2}(1-\beta)+O(\xi^2)\,.
\end{eqnarray}


\end{document}